\documentclass[pre,twocolumn,superscriptaddress,amsmath,amssymb,showpacs]{revtex4}

\usepackage{graphicx,amsfonts,amsmath}

\DeclareMathOperator{\tr}{Tr}
\DeclareMathOperator{\re}{Re}
\DeclareMathOperator{\im}{Im}

\begin{document}

\author{Katarzyna Roszak}
\affiliation{Institute of Physics, Wroc{\l}aw University of Technology, 50-370 Wroc{\l}aw, Poland}
\affiliation{Department of Condensed Matter Physics,
             Faculty of Mathematics and Physics, Charles University,
             12116 Prague, Czech Republic}
             
\author{Tom\'{a}\v{s} Novotn\'{y}}
\email{tno@karlov.mff.cuni.cz}
\affiliation{Department of Condensed Matter Physics,
             Faculty of Mathematics and Physics, Charles University,
             12116 Prague, Czech Republic}

\title{Non-Markovian effects at the Fermi-edge singularity in quantum dots}

\begin{abstract}
We study electronic transport through a quantum dot in the Fermi-edge singularity regime,
placing emphasis on its non-Markovian attributes. These are quantified by the behavior of current noise as well as trace-distance-based measure of non-Markovianity 
and found to be pronounced at low temperatures where the interplay of many-electron correlations and quantum coherence
present in the system leads to significant quantum memory effects and non-Markovian dynamics. 
\end{abstract}

\date{\today}
\maketitle

\section{Introduction}
The Fermi-edge singularity (FES) is a phenomenon originating from the interaction of 
conduction electrons with localized perturbations and is characterized by a power-law
divergence. It was first predicted theoretically
for X-ray absorption in metals \cite{Mahan:PR67} and later verified experimentally \cite{citrin73,citrin77}.
The theory then developed \cite{Mahan:PR67,Nozieres:PR69,anderson69,Ohtaka:RMP90,mahan00} has been used to describe
other situations involving a similar Hamiltonian and leading to the same power-law divergence,
such as resonant tunneling through localized levels \cite{Matveev:PRB92}. 

The FES in transport through quantum dots (QDs) 
occurs in the regime where the energy of the localised QD level(s) is comparable
to the Fermi energy of one of the leads \cite{Matveev:PRB92}. The change of occupation in the local level(s) during the tunneling 
process leads to sudden changes in the scattering potential and hence, 
to singular behavior of the current through the QD 
at resonance and power-law dependence of the current away from resonance \cite{Matveev:PRB92,Hapke-Wurst:PRB00,Frahm:PRB06,Maire:PRB07,Molenkamp:APL08}.
As shown in recent experiments \cite{Maire:PRB07,Ubbelohde:SciRep12}, current noise at such a FES displays 
characteristic behavior, including a deep minimum and a slight super-Poissonian peak, which cannot be accounted for by the Markovian theory.
Interplay of many-body correlations and quantum coherence is responsible for the large non-Markovian corrections in the FES singularity region \cite{Ubbelohde:SciRep12}. 

This paper has two objectives: first, we wish to elaborate on the theoretical method used for the explanation of the experiment in Ref.~\cite{Ubbelohde:SciRep12} and, furthermore, we explicitly demonstrate the presence of quantum memory by the recently developed measure of the quantum non-Markovianity based on the trace-distance \cite{Breuer:PRL09,Piilo:NatPhys11}. It is organized as follows. In the next section the system under study is introduced, the Hamiltonian governing its evolution is described, including the scattering term
which is responsible for the FES behavior, and the equation of motion for the reduced density matrix is formulated. In Sec. \ref{secrates}, the method for calculating
the tunneling rates is outlined which allows the non-Markovian memory kernel to be found.
Sec. \ref{secnoise} contains the formulas for the mean current and non-Markovian noise, as well as
the results. These include a comparison between Markovian and non-Markovian effects,
a description of parameter-dependences of both current and noise, and remarks on fitting experimental data. Finally, next Sec.~\ref{measure} explicitly shows the non-Markovianity of the time evolution by using the trace-distance-based measure of memory. Sec. \ref{secconc} then concludes the paper.

\section{The system and its equation of motion}
The system under study inspired by the experiments \cite{Maire:PRB07,Ubbelohde:SciRep12} consists of a quantum dot with single spin-less level 
embedded between two leads of non-interacting spin-less electrons, meaning that there is a single
channel of transport responsible for the current between
the emitter and collector leads. This may be easily extended
to the spin-full case and/or a multi-level QD and resulting multi-channel transport.
The FES occurs on the emitter lead
(henceforth referred to
as the left lead) with much smaller tunneling rate, even
at the FES threshold voltage, than the
collector (or right) lead, which means that 
tunneling from the emitter is much less probable than tunneling out of the QD
into the collector. The chemical potential of the collector is well below the QD level, while the emitter's Fermi energy can be tuned in the resonance with the level. For a schematic of the setup, we refer the reader to Fig.~1 of Ref.~\cite{Ubbelohde:SciRep12}.  

The Hamiltonian of the system is
$
H=H_{QD}+H_{L0}+H_{R0}+H_{LT}+H_{RT}+H_{X},
$
where $H_{QD}$ is the QD Hamiltonian
\begin{equation}
H_{QD}=\epsilon_0 d^{\dagger} d,
\end{equation}
with $\epsilon_0$ denoting the energy of the single level in the
QD and $d^{\dagger}, d$ being the creation and annihilation
operators of an electron in the dot. 
$H_{\alpha 0}$ with
$\alpha=L,R $ are the Hamiltonians of the left/right leads
respectively,
\begin{equation}
H_{\alpha 0}=\sum_{k_{\alpha}}\epsilon_{k_{\alpha}}c_{k_{\alpha}}^{\dagger}c_{k_{\alpha}},
\end{equation}
where the creation and annihilation operators for electrons in the left/right leads
are denoted by $c_{k_{\alpha}}^{\dagger}$, $c_{k_{\alpha}}$ and the corresponding
energies are $\epsilon_{k_{\alpha}}$. 
Tunneling from the left lead into the QD and to the right lead is governed by the Hamiltonians
$H_{\alpha T}$ ($\alpha=L,R$),
\begin{equation}
\label{tunham} H_{\alpha
T}=\sum_{k_{\alpha}}t_{k_{\alpha}}(c_{k_{\alpha}}^{\dagger}d+d^{\dagger}c_{k_{\alpha}}).
\end{equation}
Lastly, the Fermi edge singularity occurs due to the interaction of the electrons in the left
lead with the electron in the QD; this interaction is described by the scattering Hamiltonian $H_{X}$,
\begin{equation}
\label{hx}
H_{X}=\sum_{k_{L},k'_{L}}V_{k_{L},k'_{L}}c_{k_{L}}^{\dagger}c_{k'_{L}}d^{\dagger}d\equiv V_{X}d^{\dagger}d.
\end{equation}

In this system, couplings of the left and right leads to the QD state have vastly different nature and are, therefore, described by different methods. 
The right lead is not affected by the FES Hamiltonian, the lead is effectively empty (very low chemical potential) and under the experimentally well-justified assumption of constant tunnel rate  $\gamma_R=2\pi\sum_{k_{R}}|t_{k_{R}}|^2 \delta(\epsilon-\epsilon_{k_R})$ (we use units in which $\hbar=1$ throughout this paper) the transport from the QD to the right lead may be treated exactly following Ref.~\cite{Gurvitz:PRB96}. 

This allows to write the equation of motion for the evolution of the density matrix of
the QD and the left lead in a mixed form analogous to the description of charge transport through a Coulomb blockaded
double quantum dot in a dissipative environment in Refs.~\cite{Flindt:PRL08,Flindt:PRB10}.
The equation reads
\begin{equation}
\label{evolution}
\frac{d\widehat{\sigma}(t)}{dt}=\left(
\begin{matrix}
0&\gamma_R&0&0\\
0&-\gamma_R&0&0\\
0&0&-\gamma_R/2&0\\
0&0&0&-\gamma_R/2
\end{matrix}\right)\widehat{\sigma}(t)-i[H_L,\widehat{\sigma}(t)],
\end{equation}
where the first term on the right-hand side describes the effect of the traced-out
right lead \cite{Gurvitz:PRB96}, while the second term accounts for the evolution resulting from the 
rest of the Hamiltonian, i.e.~$H_{L}=H_{QD}+H_{L0}+H_{LT}+H_{X}.$ The joint QD and left lead density matrix is written in the vector form
in terms of the QD states ($|0\rangle$ and $|1\rangle$ denote the empty and occupied QD, respectively) 
$\widehat{\sigma}=\left(\begin{matrix}
\widehat{\sigma}_{00},\widehat{\sigma}_{11},\widehat{\sigma}_{01},\widehat{\sigma}_{10}
\end{matrix}\right)^T$ .

Following closely the procedure of Ref.~\cite{Flindt:PRB10}, the next step is separating Eq.~(\ref{evolution}) into four equations 
for the four elements of $\widehat{\sigma}$.
Tracing out the electron states of the left lead in the two equations
for the evolution of the diagonal elements $\widehat{\sigma}_{jj}$ ($j=0,1$),
allows us to find the evolution equations for the QD occupations
($p_0(t)$ for the empty dot and $p_1(t)$ for an electron in the dot),
\begin{eqnarray}
\label{rho11}
\frac{d p_1(t)}{dt}&=&-\frac{d p_0(t)}{dt},\\
\nonumber
\frac{d p_0(t)}{dt}&=&\gamma_R p_1(t)
-2\sum_{k_{L}}t_{k_{L}}\im\left[\tr_L\left(c_{k_{L}}\widehat{\sigma}_{01}(t)\right)\right].
\end{eqnarray}
The remaining two equations (hermitian conjugates of each other)
govern the evolution of $\widehat{\sigma}_{01}$ (and $\widehat{\sigma}_{10}$) entering the second of Eq. (\ref{rho11}) and couple to the diagonal elements $\widehat{\sigma}_{jj}$. To close the equations for the QD occupations $p_{j}(t)$ we perform physically-motivated perturbative decoupling (in relatively small $t_{k_{L}}$) of
the diagonal elements of the density matrix into $\widehat{\sigma}_{jj}(t)=p_{j}(t)\otimes\widehat{R}_{j}$, where $\widehat{R}_{j}=\exp\big(-[H_{L0}+jV_{X}-\mu_{L}\sum_{k_{L}}c_{k_{L}}^{\dagger}c_{k_{L}}]/k_{B}T\big)/Z_{J}$ are the grand-canonical density matrices of the left lead at equilibrium (described by the chemical potential $\mu_L$ and temperature $T$) when the dot is empty ($j=0$) or occupied ($j=1$). Now we find
\begin{widetext}
\begin{equation}
\begin{split}
\label{im}
\im\left[\tr_L\left(c_{k_{L}}\widehat{\sigma}_{01}(t)\right)\right]&=
\im\left[e^{-\big(\frac{\gamma_R}{2}-i\epsilon_0\big)t}\tr_L\left(e^{i(H_{L0}+V_{X})t}c_{k_{L}}e^{-iH_{L0}t}\widehat{\sigma}_{01}(0)
\right)\right]\\
&+\int_0^t dt'
p_0(t')
\re\left[e^{-\big(\frac{\gamma_R}{2}-i\epsilon_0\big)(t-t')}
\sum_{k'_{L}}t_{k'_{L}}\tr_L\left[c_{k'_{L}}^{\dagger}e^{i(H_{L0}+V_{X})(t-t')}c_{k_{L}}
e^{-iH_{L0}(t-t')}
\widehat{R}_{0}\right]\right]\\
&-\int_0^t dt'
p_1(t')
\re\left[e^{-\big(\frac{\gamma_R}{2}-i\epsilon_0\big)(t-t')}
\sum_{k'_{L}}t_{k'_{L}}\tr_L\left[e^{i(H_{L0}+V_{X})(t-t')}c_{k_{L}}
e^{-iH_{L0}(t-t')}
c_{k'_{L}}^{\dagger}\widehat{R}_{1}\right]
\right],
\end{split}
\end{equation}
which, after inserting into Eq. (\ref{rho11}) and substituting $t'\to t-t'$, leads to the non-Markovian equation of motion for the occupations
\begin{equation}\label{GME}
\begin{split}
\frac{dp_0(t)}{dt}&=-\int_0^t
dt'p_0(t-t')2\re\!\left[e^{-\big(\frac{\gamma_R}{2}-i\epsilon_0\big)t'}
G_{0}(t')\right]+\gamma_R\, p_1(t)+\int_0^t dt'p_1(t-t')2\re\!\left[e^{-\big(\frac{\gamma_R}{2}-i\epsilon_0\big)t'}
G_{1}(t')\right] + \Upsilon(t)\\
&\equiv-\int_0^t dt'\gamma_{L}(t')\,p_0(t-t') +\gamma_R\, p_1(t)+\int_0^t dt'\gamma_{L}^{b}(t')\,p_1(t-t')+\Upsilon(t)
\end{split}
\end{equation}
containing the initial-condition-dependent inhomogeneous term 
\begin{equation}
\Upsilon(t)=-2\sum_{k_{L}}t_{k_{L}}\im\left[e^{-\big(\tfrac{\gamma_R}{2}-i\epsilon_0\big)t}\tr_L\Big(e^{i(H_{L0}+V_{X})t}c_{k_{L}}e^{-iH_{L0}t}\widehat{\sigma}_{01}(0)
\Big)\right] 
\end{equation}
and the FES Green functions defined as
\begin{equation}\label{GF}
\begin{split}
G_{0}(t)&=\sum_{k_{L},k'_{L}}t_{k_{L}}t_{k'_{L}}\tr_L\left[c_{k'_{L}}^{\dagger}e^{i(H_{L0}+V_{X})t}c_{k_{L}}
e^{-iH_{L0}t}
\widehat{R}_{0}\right],\\
G_{1}(t)&=
\sum_{k_{L},k'_{L}}t_{k_{L}}t_{k'_{L}}\tr_L\left[e^{i(H_{L0}+V_{X})t}c_{k_{L}}
e^{-iH_{L0}t}
c_{k'_{L}}^{\dagger}\widehat{R}_{1}\right].
\end{split}
\end{equation}
\end{widetext}

\section{Evaluation of the memory kernel}
\label{secrates}

The evaluation of the Green functions \eqref{GF} is the central topic
of the FES theory \cite{Levitov:PRL05,Nozieres:PR69,anderson69,mahan00}
(for a review see Ref.~\cite{Ohtaka:RMP90}).
The standard procedure is to treat the Green functions
in terms of open line $L_j(t)$ and closed loop $D_{j}(t)$ factors, such that $G_j(t)=L_j(t)D_{j}(t)$. Following
Ref.~\cite{Nozieres:PR69}, these terms are found to be
\begin{equation}\label{L}
\begin{split}
D_{0/1}(t)&=(\mp it \xi_0)^{-\delta^2/\pi^2}\\
L_{0}(t)&= \sum_{k_{L}}e^{-i \epsilon_{k_{L}} t}\frac{t_{k_{L}}^2n_F(\epsilon_{k_{L}})(-\xi_0)^{2\delta/ \pi}}{(\epsilon_{k_{L}}-\mu_L)^{2\delta/ \pi}},\\
L_{1}(t)&=\sum_{k_{L}}e^{-i \epsilon_{k_{L}} t}\frac{t_{k_{L}}^2[1-n_F(\epsilon_{k_{L}})]\,\xi_0^{2\delta/ \pi}}{(\epsilon_{k_{L}}-\mu_L)^{2\delta/ \pi}},
\end{split}
\end{equation}
with $\delta$ being the phase shift associated with the scattering effects of the occupied QD level described by $V_X$ \cite{Matveev:PRB92}, 
$\xi_0$ the ultraviolet cutoff, and $n_F(\epsilon)$ denoting the Fermi-Dirac distribution of the left lead.

At zero temperature, the Fermi distribution reduces to $n_F(\epsilon)=\Theta(\mu_L-\epsilon)$,
where $\Theta(x)$ is the Heaviside step function. This considerably simplifies the problem of 
finding the functions $L_j(t)$ and leads to the known 
time-dependence \cite{Nozieres:PR69,Mahan:PR67} of the
zero-temperature FES Green functions 
\begin{equation}\label{greens}
G_{0/1}(t)=\frac{\gamma_{L}^{0}}{2\pi}\xi_{0}\Gamma\!\left(1-\frac{2\delta}{\pi}\right)e^{- i t\mu_L}(\mp it \xi_0)^{\alpha-1}.
\end{equation}
Here, $\alpha=\frac{2\delta}{\pi}-\frac{\delta^2}{\pi^2}$ is the FES critical exponent (governing the power-law dependence 
typical for the FES), $\gamma_{L}^{0}=2\pi\sum_{k_{L}}|t_{k_{L}}|^2 \delta(\epsilon-\epsilon_{k_L})$ is the non-interacting tunnel rate to the left lead stemming from the Hamiltonian \eqref{tunham}, and $\Gamma(x)$ denotes the Euler gamma function.

Knowing the Green functions, we also know the non-Markovian rates $\gamma_{L}^{(b)}(t)=2\re\Big[e^{-\big(\tfrac{\gamma_R}{2}-i\epsilon_0\big)t}
G_{0(1)}(t)\Big]$. Their Laplace transforms needed for the evaluation of the stationary current and noise read
\begin{equation}
\label{gammaz}
\begin{split}
\gamma_L(z;\Delta)&=\frac{\gamma_{L}^{0}}{\pi} \Gamma\!\left(1-\frac{2\delta}{\pi}\right)\Gamma(\alpha)\xi_0^{\alpha}\\
&\times\im\left[\left(\frac{i}{z+\frac{\gamma_R}{2}(1-i\Delta)}\right)^{\alpha} \right],\\
\gamma_L^b(z;\Delta)&=\gamma_L(z;-\Delta),
\end{split}
\end{equation}
being parametrized by the dimensionless energy-distance from the FES edge 
$\Delta=\tfrac{\mu_L-\epsilon_0}{\gamma_R/2}$.

The finite temperature result is obtained 
(as shown in Refs.~\cite{anderson69,Ohtaka:PRB84,Ohtaka:RMP90}) by substituting time $t$ in Eq.~(\ref{greens})
by the temperature-dependent expression $\frac{\sinh( \pi k_B T t)}{\pi k_B T}$.
The tunneling rate in the Laplace space $\gamma_L(z;\Delta)$ is then equal to
\begin{equation}\label{gammatemp}
\begin{split}
\gamma_L(z;\Delta)&=\frac{\gamma_{L}^{0}}{\pi}\Gamma\!\left(1-\frac{2\delta}{\pi}\right)\xi_0^{\alpha}
\im\left[\left(\frac{i}{2\pi k_B T}\right)^{\alpha}\right.\\
&\times \left. B\left(\frac{1-\alpha}{2}+\frac{z+\frac{\gamma_R}{2}(1-i\Delta)}{2\pi k_B T},\alpha\right)\right],
\end{split}
\end{equation}
where $B(x,y)$ is the Beta function.
The second of Eq. (\ref{gammaz}) for the back-flow tunneling rate $\gamma_L^b(z;\Delta)$ carries over to the finite-temperature case.

The non-Markovian generalized master equation (GME) for QD occupations
${\bf p}=(p_0,p_1)^T$ stemming from Eqs.~\eqref{GME} and \eqref{rho11} can now be rewritten in the matrix form
\begin{equation}\label{GMEtime}
\frac{d{\bf p}(t)}{dt}=\int_{0}^{t}dt'\mathbf{W} (t'){\bf p}(t-t')+{\bf\Upsilon}(t),
\end{equation}
with the memory kernel 
\begin{equation}
\label{kerneltime}
\mathbf{W}(t)=\left(
\begin{array}{cc}
-\gamma_L(t) &\gamma_L^b(t) + \gamma_R\\
\gamma_L(t) &-\gamma_L^b(t) -\gamma_R
\end{array}
\right)
\end{equation}
and the inhomogeneity ${\bf\Upsilon}(t)=(\Upsilon(t),-\Upsilon(t))^{T}$. The explicit form of this initial-condition-dependent term describing the effect of initial correlations in the system is not necessary as the stationary properties (mean current and noise) do not depend on it, since it vanishes fast enough for large times. It also drops out from the expression for the trace distance, see below in Sec.~\ref{measure}.   

For evaluation of the current and noise we need to extend the GME by including the counting field(s) and express it in the Laplace space \cite{Flindt:PRL08,Flindt:PRB10} as 
\begin{equation}
z{\bf p}(\{\chi\},z)-{\bf p}(\{\chi\},t=0)=\mathbf{W} (\{\chi\},z){\bf p}(\{\chi\},z)+{\bf\Upsilon}(\{\chi\},z),
\end{equation}
with the modified kernel (for details of introduction of the counting fields see Refs.~\cite{Flindt:PRB10,Ubbelohde:SciRep12}) 
\begin{equation}
\label{kernel}
\mathbf{W}(\{\chi\},z)=\left(
\begin{array}{cc}
-\gamma_L(z) &\gamma_L^b(z)e^{-i\chi_{L}} + \gamma_R e^{i\chi_{R}}\\
\gamma_L(z)e^{i\chi_{L}} &-\gamma_L^b(z) -\gamma_R
\end{array}
\right),
\end{equation}
where $\{\chi\}$ denotes the pair of counting fields $\chi_{L},\,\chi_{R}$ at the left/right junction respectively.

\section{Current and non-Markovian noise\label{secnoise}}
Following the scheme for calculating zero-frequency cumulants
when given the Laplace transformed memory kernel of a generalized master equation 
described in Refs.~\cite{Flindt:PRB10,Flindt:PRL08}, the mean current and
non-Markovian noise in the FES regime can be found.
The kernel is given by Eq. (\ref{kernel}) and the resulting 
stationary current reads
\begin{equation}
\label{i}
I(\Delta)=e\frac{\gamma_R\gamma_L(0;\Delta)}{\gamma_R+\gamma_L(0;\Delta)+\gamma_L^b(0;\Delta)}.
\end{equation}
The non-Markovian Fano factor is given by
\begin{equation}
\begin{split}
\label{F}
F(\Delta)&=1-\frac{2\gamma_R\gamma_L(0;\Delta)}{(\gamma_R+\gamma_L(0;\Delta)+\gamma_L^b(0;\Delta))^2}\\
&+2\gamma_R\frac{(\gamma_R+\gamma_L^b(0;\Delta))\gamma_L^{\prime}(0;\Delta)-\gamma_L^{b\prime}(0;\Delta)\gamma_L(0;\Delta)}{(\gamma_R+\gamma_L(0;\Delta)+\gamma_L^b(0;\Delta))^2},
\end{split}
\end{equation}
where the prime denotes the $z$-derivative at $z=0$, i.e.~$\left.\gamma_L^{\prime(b)}(0;\Delta)=\frac{d}{dz}\gamma_L^{(b)}(z;\Delta)\right|_{z=0}$.
The Markovian noise is given by the part of the expression without the derivatives, i.e. the first line of Eq.~\eqref{F} only.
If, additionally, no back-flow is taken into account $\gamma_L^b(0;\Delta)\to0$, Eq. (\ref{F})
reduces to the widely used Markovian expression \cite{Maire:PRB07}.

\begin{figure}
\includegraphics[width=0.5\textwidth]{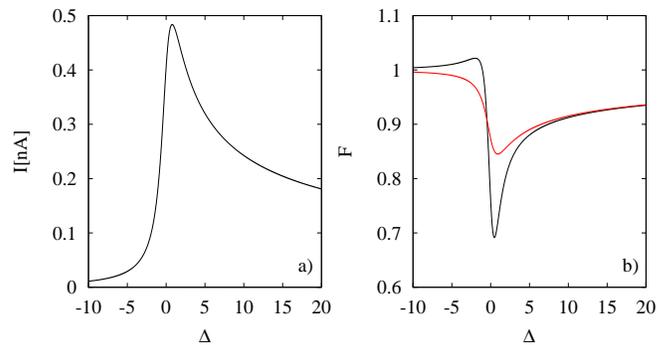}
\caption{\label{fig1} a) Zero-temperature current as a function of
$\Delta$. b) Non-Markovian (black) and Markovian (red) Fano factor at zero temperature as a function of $\Delta$.}
\end{figure}

In the following, the parameters $\alpha=0.45$ and $\gamma_R=3.4\cdot
10^{10}$ s$^{-1}$ measured in Ref.~\cite{Maire:PRB07} are used, unless explicitly
stated otherwise. The $\alpha$-dependent
prefactor $\gamma_L^0\Gamma\!\left(1-\frac{2\delta}{\pi}\right)\xi_0^{\alpha}/\pi$
can be used to control the height of the current peak; in Figs.~\ref{fig1},
\ref{fig1temp}, \ref{fig1vs}, and \ref{fig2} it is taken
such that $\gamma_{L}(0,0)/\gamma_{R}=0.09$ at zero temperature. 

In Fig.~\ref{fig1} the current and the Fano factor near the Fermi
edge singularity at zero-temperature are depicted as a function of the distance in
energy from the singularity which is represented by the parameter
$\Delta$. Hence, the energy distance is given in units of half of the tunneling rate
from the QD into the right (collector) lead. This dimensionless parameter is a more convenient theoretical measure of the energy
difference than the applied voltage $V_{SD}$, since the relation of the actual change of the energy difference $\mu_L-\epsilon_0$
to the change in applied voltage involves a leverage factor \cite{Kiesslich:PRL07,Ubbelohde:SciRep12} $\eta=(\mu_L-\epsilon_0)/e(V_{SD}-V_0)$ (where $V_0$ is the FES threshold voltage) 
which is often hard to determine experimentally.

The current shows a typical FES behavior (Fig.~\ref{fig1}a)), i.e.~sharp growth
on the low-energy side of the singularity and power-law decay on the high-energy side.
In Fig.~\ref{fig1}b), the Fano factor is plotted for both the Markovian approximation (red line) and the full non-Markovian
result (black line). Comparison of the two curves shows that near
the singularity non-Markovian features cannot be neglected,
since they change the qualitative and quantitative noise features.
Firstly, there is a significant drop of the Fano factor minimum
on the high-energy side of the singularity. Secondly,
on the low-energy side a new, super-Poissonian peak appears.
Both these features can be regarded as signatures of strongly non-Markovian dynamics.

\begin{figure}
\includegraphics[width=0.5\textwidth]{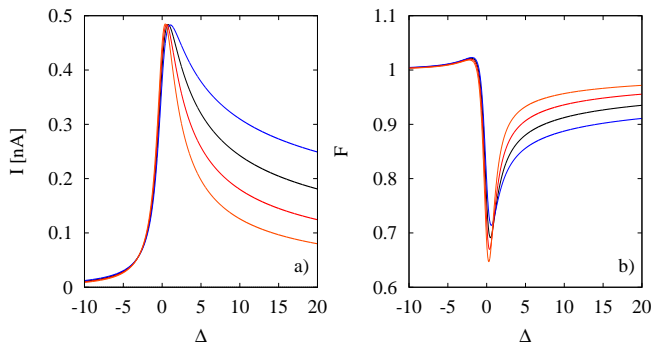}
\caption{\label{alfa} Normalized current (a) and Fano factor (b) at zero temperature as a function of
$\Delta$ for different values of $\alpha=0.45$ (black), $0.35$ (blue), $0.55$ (red), and
$0.65$ (orange).}
\end{figure}

Fig.~\ref{alfa} depicts the dependence on the critical exponent. The black curves have been transferred
from Fig.~\ref{fig1}, the blue ones correspond to $\alpha=0.35$, the red to $\alpha=0.55$
and the orange to $\alpha=0.65$.
The parameters $\gamma_L^0$ have been adjusted for the three new curves so that the 
current peaks are of the same height as for the black lines. As expected, energy dependence of the current 
on the right side of the singularity changes according to $I\propto\Delta^{-\alpha}$.
As for the Fano factor, for higher critical exponent values the minima are deeper and the 
curve on its high-energy side is steeper. Varying the critical exponent in the studied range has 
little effect on the Fano factor on the low-energy side of the singularity.

\begin{figure}
\includegraphics[width=0.5\textwidth]{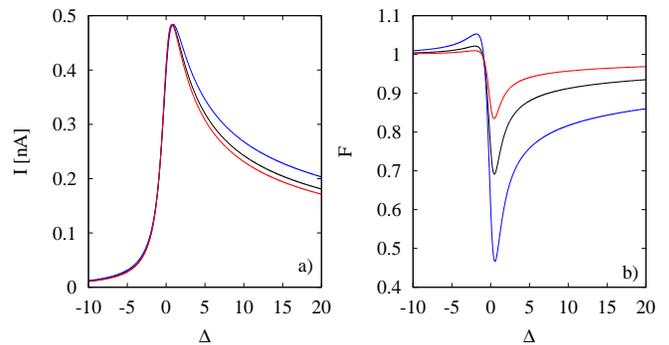}
\caption{\label{g} Normalized current (a) and Fano factor (b) at zero temperature as a function of
$\Delta$ for different values of $\gamma_R=3.4\cdot10^{10}$ s$^{-1}$ (black), $1.7\cdot
10^{10}$ s$^{-1}$ (blue), and $6.8\cdot10^{10}$ s$^{-1}$ (red). Note, that the parameter $\Delta$ stands for twice the actual energy for the red curve
and half for the blue curve with respect to that of the black curve.}
\end{figure}

\begin{figure}
\includegraphics[width=0.5\textwidth]{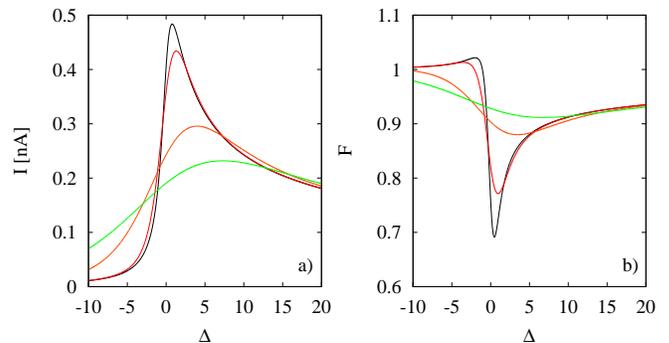}
\caption{\label{fig1temp} Current (a) and Fano factor (b) as a function of
$\Delta$ at different temperatures $T=0$ K (black) line, 0.1 K (red), 0.5 K (orange), and 1.0 K (green).}
\end{figure}

\begin{figure}
\includegraphics[width=0.5\textwidth]{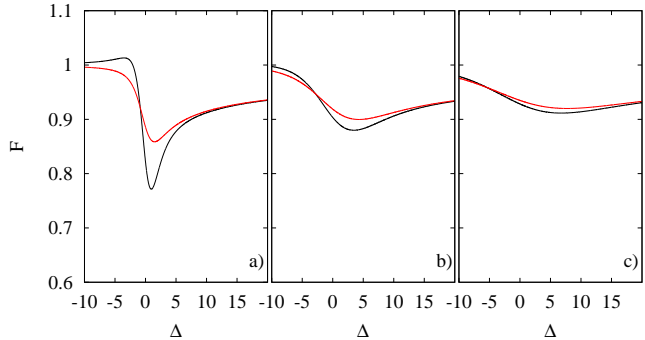}
\caption{\label{fig1vs} Markovian (red) and non-Markovian (black) Fano factor as a function of
$\Delta$ at 0.1 K (a), 0.5 K (b) and 1.0 K (c).}
\end{figure}

\begin{figure}
\includegraphics[width=0.5\textwidth]{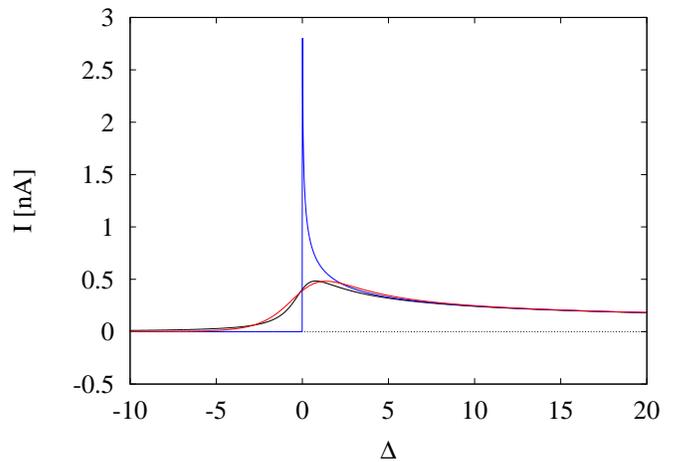}
\caption{\label{fig2} Current in the FES regime (black) at zero temperature 
compared to the approximate current \eqref{prop}
used in scaling at zero effective temperature (blue) and 0.19 K (red).}
\end{figure}

The dependence on $\gamma_R$ is shown in Fig.~\ref{g}. The black curves have been again transferred
from Fig.~\ref{fig1}, while $\gamma_R$ is halved for the blue curve and doubled for the 
red curve. The variable $\Delta$ depends on $\gamma_R$ itself,
so the actual energy changes twice as fast for the black curve
and four times as fast for the blue curve with respect to the red curve.
As in Fig.~\ref{alfa},
the parameters $\gamma_L^0$ have been adjusted in both cases so that the 
current peaks are of the same height. For the current, changing the right tunneling rate results
primarily in stretching or squeezing of the peak (in energy), while the Fano factor
undergoes large variations. Firstly, the minima and maxima are highly $\gamma_R$ dependent; the dependence
is more pronounced than the $\alpha$ dependence. Secondly, the speed at which the Fano factor 
approaches unity on the high-energy side of the minima is significantly varied. Contrarily to
Fig.~\ref{alfa}, the rise on the right side of the minimum is steeper when the minimum is 
shallower. Both of these features are important when fitting experimental data and may serve to 
check the correctness of the measured right tunneling rate $\gamma_R$.

The temperature dependence is depicted in Fig.~\ref{fig1temp}. The current peak
is damped and broadened for growing temperatures, as well as blue-shifted,
in correspondence with experimental data and theoretical predictions.
The features of the Fano factor are also damped, and
the super-Poissonian peak disappears at high temperatures. In fact, it is the 
non-Markovian features
becoming less pronounced; they lose any relevance at higher temperatures.
This can be seen in Fig.~\ref{fig1vs} where the
Markovian (red line) and non-Markovian (black line) Fano factors are compared 
for the same temperatures as in Fig.~\ref{fig1temp}.
The crossover temperature for the irrelevance of non-Markovian corrections
depends on system parameters $\alpha$ and $\gamma_R$, roughly being of the order of $\gamma_{R}$, see Sec.~\ref{measure}.

Finally, let us briefly comment on some consequences of our results for the comparison with experiments. One of the methods used for finding the critical exponent $\alpha$ 
relies on collapsing of the current curves corresponding to different temperatures 
to a common curve \cite{Molenkamp:APL08,Frahm:PRB06}. This is done by assuming the proportionality
\begin{equation}
\label{prop}
I\propto
\im\left[\left(\frac{i}{2\pi k_B T}\right)^{\alpha}
B\left[\frac{1-\alpha}{2}-\frac{i \eta(V_{SD}-V_0)e}{2\pi k_B T},\alpha\right]\right],
\end{equation}
where $V_{SD}$ is the source-drain voltage responsible for the QD level energy shift, 
$V_0$ is the FES threshold voltage corresponding to the resonance of the QD level with the left lead's chemical potential
$\mu_L$
and $\eta$ is the leverage factor, such that $\eta(V_{SD}-V_0)e=\Delta\frac{\gamma_R}{2}$.
If Eq. (\ref{prop}) correctly describes the measured $I-V$ curves then
rescaling the current 
$I\rightarrow I_T=IT^{\alpha}$ and the source-drain voltage
$V_{SD}\rightarrow V_T=(V_{SD}-V_0)e/k_BT$ which lifts any temperature dependence from it,
\begin{equation}
I_T\propto
\im\left[\left(\frac{i}{2\pi k_B}\right)^{\alpha}
B\left[\frac{1-\alpha}{2}-\frac{i \eta V_T}{2\pi},\alpha\right]\right],
\end{equation}
will reduce all temperature-dependent curves to one
for the correct critical exponent.

Eq.~(\ref{prop}) is a twofold approximation. Firstly, the current, Eq.~(\ref{i}),
is a non-trivial function of the tunneling rates, for which the proportionality 
should hold. It reduces to $I\approx e\gamma_L(0;\Delta)$ for $\gamma_R\gg \gamma_L(0;\Delta)$;
the theoretical analysis presented in this paper requires this condition to be fulfilled,
nonetheless, small deviations of the current from the left tunneling rate should be expected 
near the singularity. 

Secondly and more importantly, comparing Eq.~(\ref{prop}) with Eq.~(\ref{gammatemp}) shows an additional 
real, temperature dependent factor in the Beta function
$ \gamma_R/(4\pi k_B T)$. 
This factor stems from the first term on the right hand side of Eq. (\ref{evolution}) which is
responsible for the effect of the right lead on the QD. It describes the damping of the 
coherence of the QD state and is typically neglected.
Although this factor is negligible away from the singularity,
it plays a significant role in reducing the FES peak. The discrepancy is usually handled
by introducing an effective temperature $k_BT^*=\sqrt{(k_BT)^2+\Gamma_i^2}$,
where the temperature enhancement is wrongly attributed to a non-zero 
intrinsic linewidth of the QD state $\Gamma_i$
ascribed to the hybridization of the QD electron state with the emitter (left) lead states.

The current at zero temperature as described by Eq.~(\ref{i}) is compared with the corresponding curve
given by Eq.~(\ref{prop}) at 0 K and 0.19 K in Fig.~\ref{fig2}.
The effective-temperature treatment reproduces the current curves given by Eqs.~(\ref{i})
and (\ref{gammatemp}) at relatively high temperatures; it is then possible to 
collapse them to a common curve and find the critical exponent $\alpha$ with reasonable
precision (if the leverage factor $\eta$ is known). The temperature limit of its applicability
depends on the critical exponent and the tunneling rate of the right lead $\gamma_R$
and is of the order of $0.1$ K. At very low temperatures, collapsing different temperature
curves does not yield the correct critical exponent.

\section{Direct measure of non-Markovianity of the time evolution}\label{measure}

In this final section we address the non-Markovianity of the evolution directly in the time domain using the recently developed \cite{Breuer:PRL09} measure of quantum memory based on the trace distance of two initial conditions, which was successfully experimentally tested in a designed quantum-optical experiment \cite{Piilo:NatPhys11}. The measure is constructed as follows: first, for any two  density matrices $\rho_{1,2}$ their trace distance is defined as $D(\rho_{1},\rho_{2})=\tfrac{1}{2}\tr|\rho_{1}-\rho_{2}|$ with $|A|\equiv \sqrt{A^{\dagger}A}$. In our case, we only deal with diagonal elements of the density matrix $\rho=\mathrm{diag}(p_{0},p_{1})$, which together with the normalization condition $p_{0}+p_{1}=1$ yields for two probability distributions $\mathbf{p}_{1,2}$   
\begin{equation}
D(\mathbf{p}_{1},\mathbf{p}_{2})=\frac{1}{2}\big(|p_{0}^{1}-p_{0}^{2}|+|p_{1}^{1}-p_{1}^{2}|\big)=|p_{0}^{1}-p_{0}^{2}|\equiv |\Sigma|.
\end{equation}
Now, the measure of non-Markovianity $\mathcal{N}$ of a dynamical map inducing time evolution (here basically the evolution kernel $\mathbf{W}(t)$, Eq.~\eqref{kerneltime}) is defined as \cite{Breuer:PRL09,Piilo:NatPhys11} $\mathcal{N}(\mathbf{W})=\mathrm{max}_{\mathbf{p}_{1,2}(0)}\int_{\dot{D}>0} dt \dot{D}(\mathbf{p}_{1}(t),\mathbf{p}_{2}(t))$, i.e.~sum of stretches of the distance, where the distance is increasing, maximized over the pairs of initial conditions. The increase in the distance is a sign of back-flow of information into the system and is a hallmark of non-Markovian behavior \cite{Breuer:PRL09}.

\begin{figure}
\includegraphics[width=0.5\textwidth]{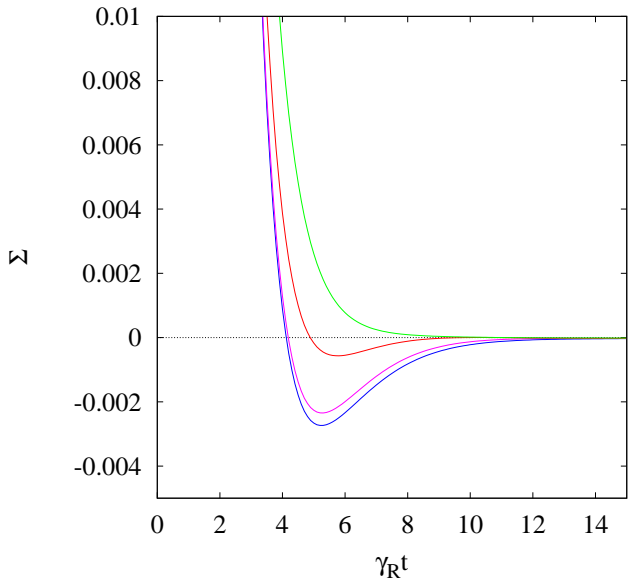}
\caption{\label{MemTemp} Memory measure $\Sigma(t)$ (see the main text for details) as a function of the dimensionless time $\gamma_{R}t$ for various values of temperature $k_{B}T/\gamma_{R}=0$ (blue), 1/10 (purple; $T\doteq0.03$ K), 1/3 (red; $T\doteq0.09$ K), and 1 (green; $T\doteq0.26$ K) at $\Delta=0$.}
\end{figure}

\begin{figure}
\includegraphics[width=0.5\textwidth]{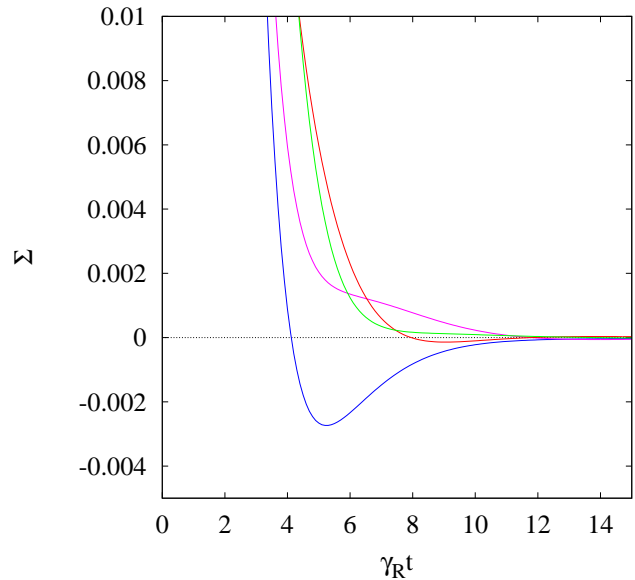}
\caption{\label{MemDelta} Memory measure $\Sigma(t)$ (see the main text for details) as a function of the dimensionless time $\gamma_{R}t$ for various values of $\Delta=0$ (blue), 1 (purple), 2 (red), and 3 (green) at zero temperature.}
\end{figure}

We can find easily the time evolution of the quantity $\Sigma(t)\equiv p_{0}^{1}(t)-p_{0}^{2}(t)$ from Eq.~\eqref{GMEtime} leading to the expression in the Laplace space
\begin{equation}\label{eq:measure}
\Sigma(z)=\frac{\Sigma(t=0)}{z-W_{11}(z)+W_{12}(z)}=\frac{\Sigma(t=0)}{z+\gamma_{R}+\gamma_{L}(z)+\gamma_{L}^{b}(z)}.
\end{equation}
Obviously, $\Sigma(t)$ (given as the backward Laplace transform of the above expression) is simply proportional to the initial condition $\Sigma(t=0)$ --- maximization over the initial condition
in the measure $\mathcal{N}(\mathbf{W})$ requires taking its maximal value $\Sigma(t=0)=1$. The measure uses in its definition the absolute value of $\Sigma(t)$, which indeed may become negative, in order to insure the positivity of the distance. Straightforward inspection of the behavior of the distance reveals that the non-Markovianity measure can be obtained in our case by the same prescription applied directly to $\Sigma$ and, in practice, this means that the non-Markovianity measure is basically equal to the depth of the $\Sigma(t)$-curve minimum, see Figs.~\ref{MemTemp}, \ref{MemDelta}.

In Fig.~\ref{MemTemp} we show the time evolution of $\Sigma(t)$ at $\Delta=0$ for varying values of temperature, while Fig.~\ref{MemDelta} depicts the $\Delta$-dependence at zero temperature. Most importantly, we see that the memory measure is generically non-zero (i.e., there is a finite dip below zero in the curves whose depth gives essentially  the memory measure) thus proving by a different method complementary to the noise the existence of the quantum memory in the system. In Fig.~\ref{MemTemp} we observe an expected monotonous decrease of the non-Markovianity measure with increasing temperature fully in line with previous findings on noise shown in Fig.~\ref{fig1vs}. In Fig.~\ref{MemDelta} there is, however, a surprising non-monotonic dependence on $\Delta$, namely, the curve for $\Delta=1$ does not exhibit any minimum and the measure assumes zero value. Moreover, Eq.~\eqref{eq:measure} contains only a symmetric-in-$\Delta$ combination of rates and the $\Sigma(t)$-curves as well as the memory measure are therefore equal for $\pm\Delta$ in disagreement with the obvious $\Delta$-asymmetry of the Fano factor curves in Fig.~\ref{fig1}b). This may be caused by the neglect of the off-diagonal matrix elements in our theory for short times (they are irrelevant in the stationary state and, thus, do not influence the noise). Their proper inclusion (though not obvious how to accomplish) might increase the measure by enlarging the space over which the maximization takes place, and potentially also break the symmetry in $\Delta$ which is now present. Despite of this issue, the global picture in Fig.~\ref{MemDelta}, i.e.~studying not just a single curve but a whole set of them, clearly reveals the presence of the quantum memory in our system.     
\newline

\section{Conclusion\label{secconc}}
We have studied electronic current transport through a quantum dot in the Fermi edge singularity regime and have shown
that the associated current noise displays pronounced non-Markovian features at low temperatures.
This is due to the interplay of many-electron correlations and quantum coherence
present in the system which leads to the significance of quantum memory effects
and strongly non-Markovian dynamics. The features include a pronounced deepening
of the minimum in the Fano factor on the high-energy side of the singularity and an appearance
of a super-Poissonian maximum on the low-energy side.

The study of the current and noise curves parameter dependence shows
that although changing the critical exponent
and the tunneling rate of the collector (right) lead have similar effects on 
the current curve, it leads to qualitatively different variations of the Fano
factors energy dependence. 
Hence, noise measurements may be used to check the correctness of the 
experimentally found values of the critical exponent governing the FES power-law
divergence. This is important since there is an ambiguity when fitting the current 
curves alone (due to problems in establishing the correct leverage factor
between energy and applied voltage)
and we have also shown that the method of collapsing different-temperature curves 
to find the critical exponent is approximate and fails at very low temperatures.

Finally, we have checked our noise findings by applying a newly developed measure of quantum non-Markovianity to our problem. The results indeed confirm the presence of quantum memory in the dynamics of the FES transport, although the application of the measure contains slight inconsistencies pointing towards inadequacy of theoretical description for short times. This could be a topic of further research alongside an obvious task of deriving the governing master equation from a systematic microscopic theory, which is an open nontrivial issue at the moment.    

\begin{acknowledgments}
We acknowledge support by the Czech Science Foundation via grant No.~204/12/0897 and the Charles University Research Center ``Physics of Condensed Matter and Functional Materials" (T.~N.) and by the TEAM programme of the Foundation for Polish Science, co-financed from the European Regional Development Fund (K.~R.).     
\end{acknowledgments}

\bibliography{FES}

\end{document}